\documentclass[aps,pra,numerical,superscriptaddress,showpacs,amsmath,amssymb,
floatfix,
preprint,
groupedaddress
]{revtex4-1}
\usepackage{graphicx}
\usepackage{dcolumn}
\usepackage{bm}
\usepackage[english]{babel}
\usepackage{siunitx}
\usepackage{nicefrac}
\begin{document}

\title{Rydberg atom-based Electrometry Using a Self-Heterodyne Frequency Comb Readout and Preparation Scheme}
\author{Katelyn Dixon, Kent Nickerson, Donald W. Booth and James P. Shaffer}

\affiliation{Quantum Valley Ideas Laboratories, 485 Wes Graham Way, Waterloo, ON N2L 6R1, Canada}

\date{\today}

\email{jshaffer@qvil.ca}

\begin{abstract}
Atom-based radio frequency electromagnetic field sensing using atomic Rydberg states is a promising technique that has recently attracted significant interest. Its unique advantages, such as extraordinary bandwidth, self-calibration and all-dielectric sensors, are a tangible improvement over antenna-based methods in applications such as test and measurement, and development of broad bandwidth receivers. Here, we demonstrate how an optical frequency comb can be used to acquire data in the Autler-Townes regime of Rydberg atom-based electrometry in a massively parallel fashion, eliminating the need for laser scanning. Two-photon electromagnetically induced transparency read-out and preparation of cesium is used for the demonstration. A flat, quasi-continuous optical comb is generated with the probe laser at 852$\,$nm using an electro-optic modulator and arbitrary waveform generator. A single frequency coupling laser at 509$\,$nm is tuned to the Rydberg launch state. An enhanced transmission signal is obtained using self-heterodyne spectroscopy.  The comb signal is beat against a local oscillator derived from the single frequency probe laser on a fast photodiode. The transmission of each probe laser comb tooth is observed. We resolve electromagnetically induced transparency peaks with linewidths below 5$\,$MHz, with and without laser locking. Radio frequency electromagnetic fields as low as 66$\,\mu$Vcm$^{-1}$ are detected with sensitivities of 2.3$\,\mu$Vcm$^{-1}$Hz$^{-1/2}$. The method offers a significant advantage for reading-out electromagnetically induced transparency and Autler-Townes splitting as neither laser needs to be scanned and slow frequency drifts can be tolerated in some applications. The method enables the detection of the amplitude of a pulsed radio frequency electromagnetic field when the incoming pulse Autler-Townes splits the electromagnetically induced transparency peak.
\end{abstract}

\maketitle

\section*{Introduction}

The detection of radio frequency (RF) electromagnetic fields is a crucial tool for applications such as radar, communications, test and measurement, environmental science, and materials characterization \cite{fan2015atom,anderson2017continuous,holloway2014sub}.
Rydberg atom-based sensing of RF electromagnetic fields has recently gained attention as a promising alternative to existing antenna-based standards, as atom-based methods provide high stability, accuracy and reproducibility \cite{sedlacek2012microwave,holloway2014broadband,anderson2016optical,gordon2014millimeter}. In contrast to antenna-based methods, Rydberg atom-based electromagnetic field sensors are capable of measurements which are self-calibrating, have a large carrier bandwidth and minimally distort the target RF electromagnetic field \cite{sedlacek2012microwave,holloway2014broadband}. The vapor cells used for sensing can be made in a wide range of shapes and sizes. Different types of materials can also be used for their construction. The vapor cells can be tailored to specific applications.

In one mode of Rydberg atom-based sensing, the Autler-Townes splitting induced by a target RF electromagnetic field in an atomic vapor cell is detected via electromagnetically induced transparency (EIT) \cite{sedlacek2012microwave}. EIT is a quantum interference phenomenon in which the probability amplitudes for transitions between quantum states interfere destructively to reduce the absorption of one of the lasers, referred to as the probe laser \cite{marangos1998electromagnetically}. The destructive interference in Rydberg atom EIT can be disrupted by inducing an Autler-Townes splitting of a Rydberg state through the use of a near-resonant RF electromagnetic field. The spectral splitting of the Rydberg state energy level produces a splitting of the EIT transmission peak that is read out by the probe laser \cite{sedlacek2012microwave}. It is this splitting of the EIT peak that enables self-calibrated operation because it depends only on the structure of the atom, through the transition dipole moment and Planck's constant. As the RF electromagnetic field amplitude is decreased, the Autler-Townes peaks gradually overlap and the change in amplitude of the EIT transmission due to the RF electromagnetic field can be used to measure the magnitude of the RF electric field. The change in EIT transmission amplitude due to the RF field decreases until the sensitivity limit is reached in this second operation mode. These methods have been used for the detection of RF electromagnetic fields down to field strengths of 1$\,\mu$V$\,$cm$^{-1}$ and with sensitivities as low as 240$\,$nV$\,$cm$^{-1}$Hz$^{-1/2}$ \cite{fan2015atom,bohaichuk2022origins,kumar17,kumar17a}. The sensitivity can be further improved to $<$55$\,$nV$\,$cm$^{- 1}$Hz$^{-1/2}$ through the use of an auxiliary RF field in a heterodyne detection scheme, but requires a conventional antenna \cite{jing2020atomic,simons19}. We note that these quoted sensitivities also depend on the strength of the transition dipole moment on the sensing transition, as well as geometric and laser-matter interaction parameters.

The standard experimental apparatus used for Rydberg atom-based sensing utilizes single-frequency probe and coupling lasers which are counter-propagated within an alkali gas vapor cell. To probe the EIT peak and RF electromagnetic field-induced splitting, the frequency of one of the lasers is scanned over the transmission region while the frequency of the other laser is fixed to the transition. To determine the sensitivity limit of the system,
the wavelengths of both lasers are locked to the center of the EIT peak and the change in transmission is measured as a function of the RF electromagnetic field strength \cite{sedlacek2012microwave}. Pulsed RF fields, with durations as low as 50$\,$ns, have also been used to examine the transient atomic response by monitoring the peak amplitude as a function of time using fixed frequency lasers \cite{bohaichuk2022origins}. Neither the laser scanning nor the pulsed detection approaches allow for real-time measurement of the full EIT transmission spectra, limiting the amount of information which can be obtained.  High-speed tuning of the laser can distort the spectral lineshape, particularly when combined with modulation techniques aimed at increasing signal-to-noise levels \cite{hebert2016real}. More importantly, for many applications the time of arrival is not known. In some time-dependent applications like radar and communications it is useful to determine the pulse amplitude, but once the EIT peaks are split the amplitude dependence of the response saturates, limiting the dynamic range of electromagnetic field strengths which can be detected.

A frequency comb is a light field consisting of an optical spectrum made up from a set of equally spaced discrete frequency lines, resembling a comb in frequency space. A number of techniques have been developed to generate frequency combs including mode-locked lasers, optical micro-resonators, nonlinear optical fibers, acousto-optic modulators and electro-optic modulators \cite{ozharar2007ultraflat,deschenes2014frequency,parriaux2020electro}. Among these, electro-optic modulators (EOMs) offer superior tunability in both tooth spacing and comb shape especially when the comb tooth spacing is required to be $<$10$\,$MHz. Electro-optic frequency combs allow the comb to be optimized for many different applications \cite{parriaux2020electro}. In our case, the electro-optic frequency comb is advantageous because a very fine comb tooth spacing can be achieved ($<$1$\,$MHz), with sufficient breadth to cover the necessary spectral bandwidth ($\sim$GHz, the spectral width of the Doppler broadened probe laser transition).

Broadband atomic spectroscopy is a promising application of optical frequency combs, as a comb allows for the instantaneous interrogation of a wide spectral range when used to directly excite or probe a sample. Several techniques for frequency comb spectroscopy have been demonstrated including direct frequency comb spectroscopy, Ramsey comb spectroscopy, frequency comb Fourier transform spectroscopy and dual-comb spectroscopy \cite{picque2019frequency}. Frequency comb interference spectroscopy is a popular technique for the interrogation of atomic transitions, in which a frequency comb is passed through a sample and then beat against a single frequency laser, down-converting the comb modes for electronic processing \cite{urabe2012absorption,deschenes2014frequency}. Frequency combs in this configuration have been used to probe various transitions in atomic gasses such as cesium and potassium, and have demonstrated time resolutions as low as 1.12$\,$$\mu$s \cite{hebert2016real,long2016multiplexed,hebert2015self}. 

In this work, we use an electro-optically generated frequency comb to detect RF electromagnetic fields by probing EIT transmission and RF electromagnetic field induced peak splitting in a cesium vapor cell. We counter-propagate a frequency comb probe centered at $\sim 852\,$nm with a fixed-frequency coupling laser, $\sim 509\,$nm, allowing us to measure a spectral bandwidth of $\sim$100$\,$MHz without needing to sweep the frequency of either laser. The probe comb is beat against a single-frequency carrier beam in a self-heterodyne spectroscopy scheme. The technique greatly simplifies the optical setup required to conduct RF sensing measurements while requiring the addition of only an EOM, an arbitrary waveform generator (AWG) and a spectrum analyzer. The simplification of the RF electromagnetic field sensing platform makes this technique amenable to a much wider range of applications than previous methods reliant on swept frequency lasers and precision wavelength measurements, such as pulse amplitude detection over a broad range of RF electromagnetic field strengths. The ability to instantaneously image the full transmission spectrum, which is not possible with swept-laser methods, provides significant additional information which is useful in the measurement of pulsed fields. We additionally demonstrate that EIT peaks can be measured when both lasers are unlocked, further simplifying the minimum optical requirements of the system for some applications.

\section*{Methods}
To generate the optical frequency comb an EOM is driven by an AWG to induce the optical phase modulation. The function generated by the AWG is a repeated chirped sinusoid of the form
\begin{equation}
    f(t)=V_0\sin \left(2\pi\left(f_0t+\frac{(f_1-f_0)f_{c}}{2}t^2\right)\right),
\end{equation}
where $V_0$ is the voltage amplitude of the modulation, $f_o$ and $f_1$ are the initial and final frequencies of the sinusoid, and $f_{c}$ is the frequency spacing between the comb teeth \cite{long2016multiplexed}.
This comb function is highly tunable, as the spectral width and tooth spacing can be set to any desired value using $f_0$, $f_1$, and $f_{c}$. The only limitations on the comb parameters are the sampling rate and the buffer size of the AWG, since these limit the comb tooth spacing, $f_c$, and the spectral width of the comb, $2(f_1-f_0)$.
To generate a finely spaced, quasi-continuous comb with sufficient frequency span, we set $V_0=10\,$V, $f_0=0\,$MHz, $f_1=50\,$MHz and $f_{c}=10\,$kHz over a time span of $t=100\,\mu$s. The corresponding optical modulation applied to the EOM is 
\begin{equation}
    A(t)=e^{i\omega_ct}e^{-i\frac{\pi f(t)}{{V_\pi}}} ,
\end{equation}
where $\omega_c$ is the carrier frequency and $V_{\pi}=2.1\,$V is the half-wave voltage of the EOM \cite{ozharar2007ultraflat}. In this work the modulation amplitude $V_0$ of $f(t)$ is significantly less than $V_\pi$. The low driving voltage leaves a strong central peak at the center of the comb, but assures fairly constant power to the other comb lines resulting in a uniform profile. The AWG used in this work restricts the time span of the modulation function to 100$\,\mu$s for the selected tooth spacing. While the duration of the modulation can be reduced by increasing the tooth spacing at fixed bandwidth, this comes at the expense of reduced spectral resolution of the EIT signal.
The optical comb used for the experiments, corresponding to the previous stated parameters, is shown in Fig~\ref{fig:comb}a and b. The comb has a frequency span of 100$\,$MHz and is composed of 10000 teeth with a power variation of 3.06$\,$dB. The experimentally measured comb bears excellent resemblance to the simulated comb, albeit with a slight change in tooth power across the comb width. 

\begin{figure}[htbp]
\centering
\includegraphics[width=0.8\textwidth]{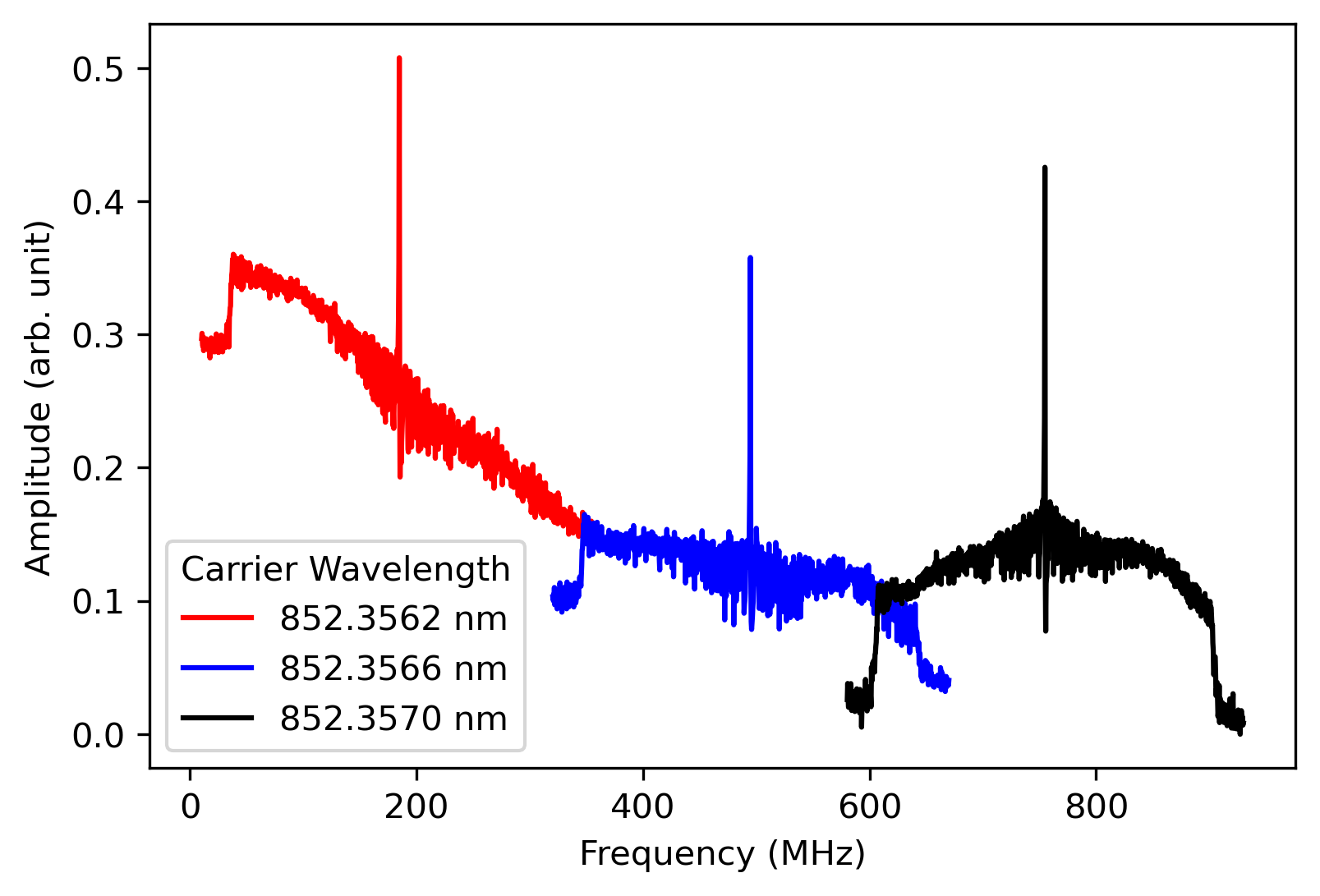}
\caption{Doppler broadened cesium D2 absorption imaged by three optical frequency combs. The comb passes through the vapor cell. As the carrier frequency of the comb is shifted it maps out the Doppler profile of the Cs D2 transition. For this data, $f_c = 0.1\,$MHz, $f_0 = 0\,$MHz and $f_1 = 150\,$MHz.}
\label{fig:doppler}
\end{figure}

After passing through the vapor cell, the comb, shown in Fig~\ref{fig:comb}b, is distorted by absorption of light on the D2 line of cesium. To demonstrate that the comb tilt observed in Figure \ref{fig:comb}a was caused by Doppler broadened absorption of the cesium D2 line \cite{im2001saturated}, the carrier frequency of the comb was shifted to image the absorption curve. The carrier wavelength of the comb was varied from 852.3562 nm to 852.3570 nm in steps of 0.0004 nm as measured by a commercial wavemeter. An image of the comb was collected at each setting, and all three are displayed in Figure \ref{fig:doppler}. The combined profile of these combs maps the Doppler broadened absorption curve characteristic of room temperature cesium at this wavelength.

\begin{figure}[htbp]
\centering
\includegraphics[width=\textwidth]{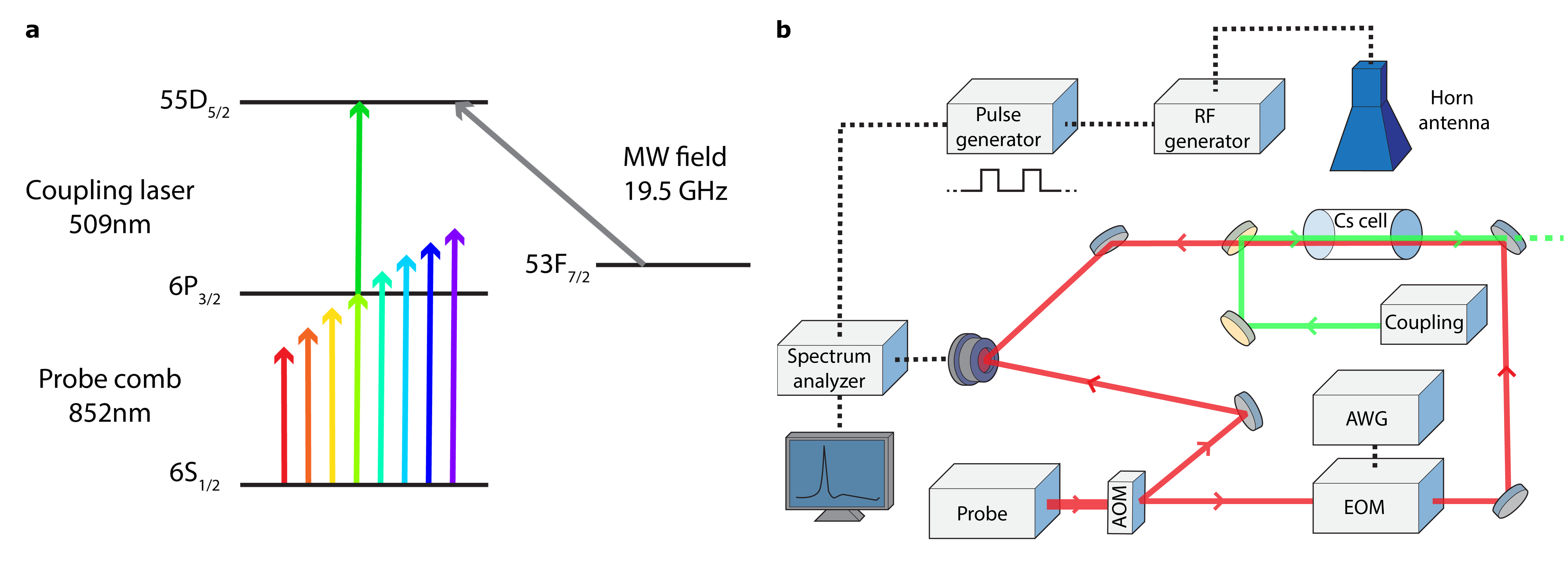}
\caption{(a) Energy level diagram illustrating the four-level system used in the measurements. Optical frequency components within the comb are illustrated in different colours. (b) Schematic of the experimental scheme. Free space probe and coupling beams are shown in red and green, respectively. Electronic connections are shown as dotted lines.}
\label{fig:shematics}
\end{figure}

Figure~\ref{fig:shematics}b illustrates the optical setup used to conduct the measurements. A probe laser with a wavelength of 852$\,$nm, resonant with the cesium D2 transition, passes through an acousto-optic modulator (AOM). A local oscillator for self-heterodyne measurements is generated by picking off the first order output, which is frequency shifted by 185$\,$MHz. The frequency comb uses the zeroth order AOM output. The probe laser comb has a wavelength of $\sim$852$\,$nm, a beam diameter of $345.6\, \pm\,1\,\mu$m and a Rabi frequency of 0.11$\,$MHz per tooth. 
The coupling laser has a wavelength of $\sim$509$\,$nm 
, a beam diameter of $260.3\, \pm\,2\,\mu$m and a Rabi frequency of 3.75$\,$MHz. The probe and coupling lasers counter-propagate through a cesium vapor cell. The room temperature vapor cell is cylindrical with a 2.54$\,$cm diameter and 40$\,$mm length. The vapor cell is constructed of sealed glass containing cesium.
Both lasers are locked using the Pound-Drever-Hall method to an ultra-stable Fabry-Perot cavity \cite{black2001introduction}. After passing through the vapor cell the probe beam is beat against the local oscillator in a self-heterodyne spectroscopic measurement using a fast photodetector with a bandwidth of 200 MHz. The beat signal is read-out using a spectrum analyzer to visualize the comb and EIT peaks. Averages of 50 analyzer scans of 200$\,$ms duration using a bandwidth of 1$\,$kHz were taken. Optimally, a real-time spectrum analyzer can be used to decrease the signal acquisition time, but only a swept-tuned spectrum analyzer was available for this experiment. Nevertheless, the results will be similar, if not identical, with a real-time spectrum analzyer. To isolate the EIT peak from the comb structure, a background signal is collected by blocking the coupling laser and eliminating the induced transmission. The coupling laser is then reintroduced, generating EIT, and the spectrum is collected once again. The background signal, which is stable in time, is subtracted from the EIT signal. The background subtraction produces an isolated EIT peak without the underlying comb structure.

To generate the target RF electromagnetic field a horn antenna driven by a RF generator 
emitting at 19.3965$\,$GHz is placed in the far-field of the cesium vapor cell. The RF field propagates along the radial direction of the cylindrical vapor cell. The drift of the signal was minimized and the sensitivity was improved for the low-power RF electromagnetic field measurements using a custom remote control program to control the spectrum analyzer. The program enables the collection of the signal and background in direct succession so that they can be subtracted. To achieve this, a pulse generator 
emitting a square wave with a frequency of 14$\,$Hz and 50\% duty cycle is used to pulse the RF signal and trigger the spectrum analyzer to collect either background (RF off) or signal (RF on) spectra. The pulse cycle is repeated 50 times, after which the background is subtracted from the EIT signal and the resulting isolated signals are averaged. The process directly measures the difference in peak amplitude induced by the RF electromagnetic field, which tends to zero as the electromagnetic field intensity reaches the sensitivity threshold of the system.

\section*{Results}
\begin{figure}[htbp]
\centering
\includegraphics[width=\textwidth]{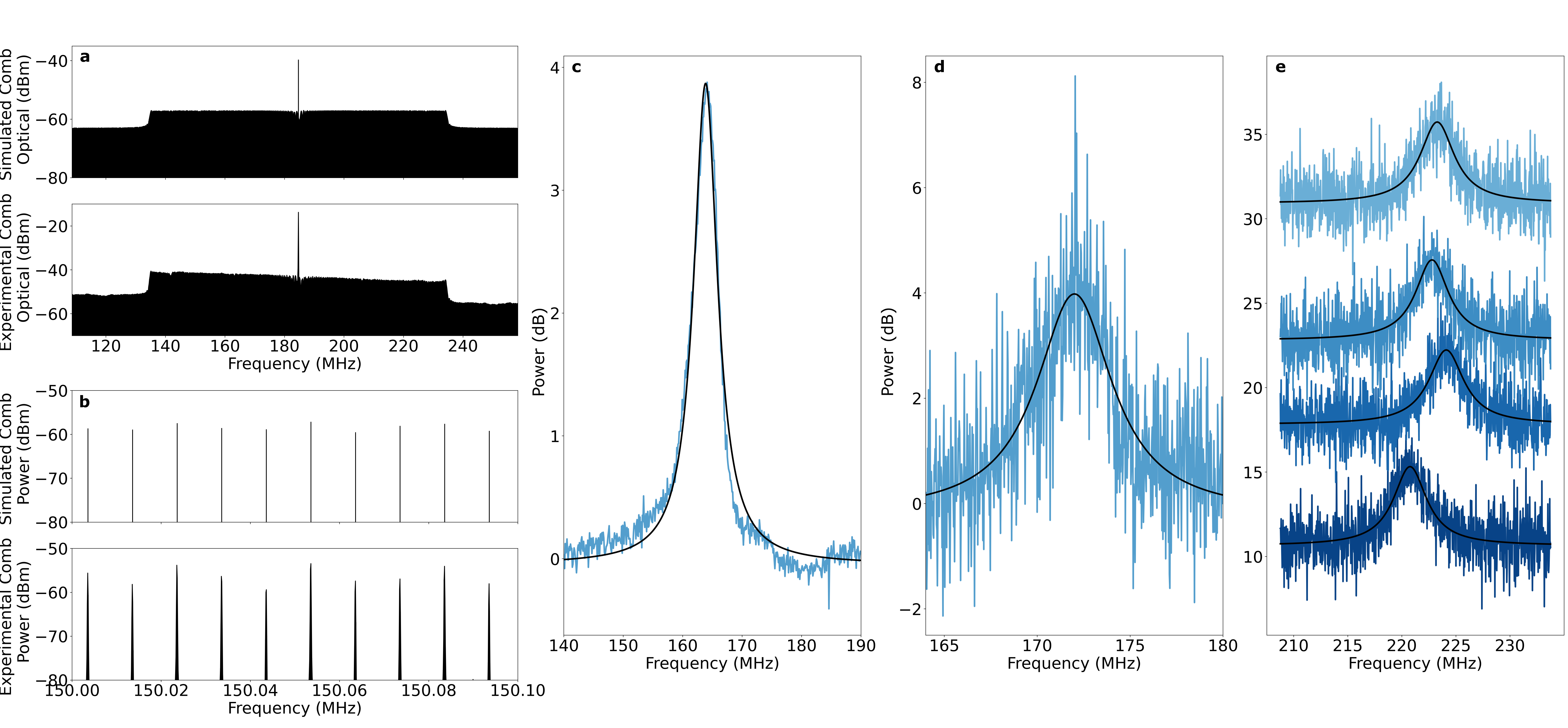}
\caption{(a) The resulting optical comb as generated by both simulation and experiment. The experimental plot shows the comb after passing through the room temperature cesium vapor cell (b) A magnified section of the comb teeth with a frequency span of 100$\,$kHz. (c-d) EIT peaks obtained with this optical comb without RF coupling. Raw data is shown in blue, and a Lorentzian fit is shown in black. (c) EIT peak obtained with both lasers locked by averaging 50 scans with a sweep time of 200$\,$ms. The linewidth is 4.9$\pm 0.1\,$MHz. (d) EIT peak obtained with both lasers unlocked by taking a single scan with a sweep time of 1$\,$ms. The linewidth is 4.7$\pm0.1\,$MHz. (e) EIT peaks collected in succession, illustrating the jitter in peak position when both lasers are unlocked.}
\label{fig:comb}
\end{figure}

The EIT signal obtained with both lasers locked is shown in Figure~\ref{fig:comb}c, after background subtraction and averaging 50 spectrum analyzer scans of 200$\,$ms duration. A Lorentzian fit of the peak is shown, which has a full-width at half-maximum of 4.9$\pm0.1\,$MHz. To illustrate the possibility of measuring EIT signals without locking the lasers, an EIT peak is collected with both the coupling and probe lasers unlocked. The peak obtained with free running lasers is shown in Figure~\ref{fig:comb}d. Due to drift in the unlocked laser wavelengths, the scan time is reduced to 1 ms and no averaging is performed. The signal-to-noise ratio is reduced, but the peak is clearly visible. The EIT peak has a full-width at half-maximum of 4.7$\pm0.1\,$MHz. The data demonstrates that the frequency comb enables the measurement of the EIT peak without laser locking. To illustrate the effects of laser drift and jitter on the EIT spectra, four EIT peaks collected in succession are shown in Figure~\ref{fig:comb}e. The ability to probe the EIT transmission spectrum without needing to lock either the probe or coupling laser is one advantage of frequency comb spectroscopy.  Not having to scan the lasers greatly simplifies the optical control systems required to perform the measurement in some use cases.

\begin{figure}[htbp]
\centering
\includegraphics[width=0.7\textwidth]{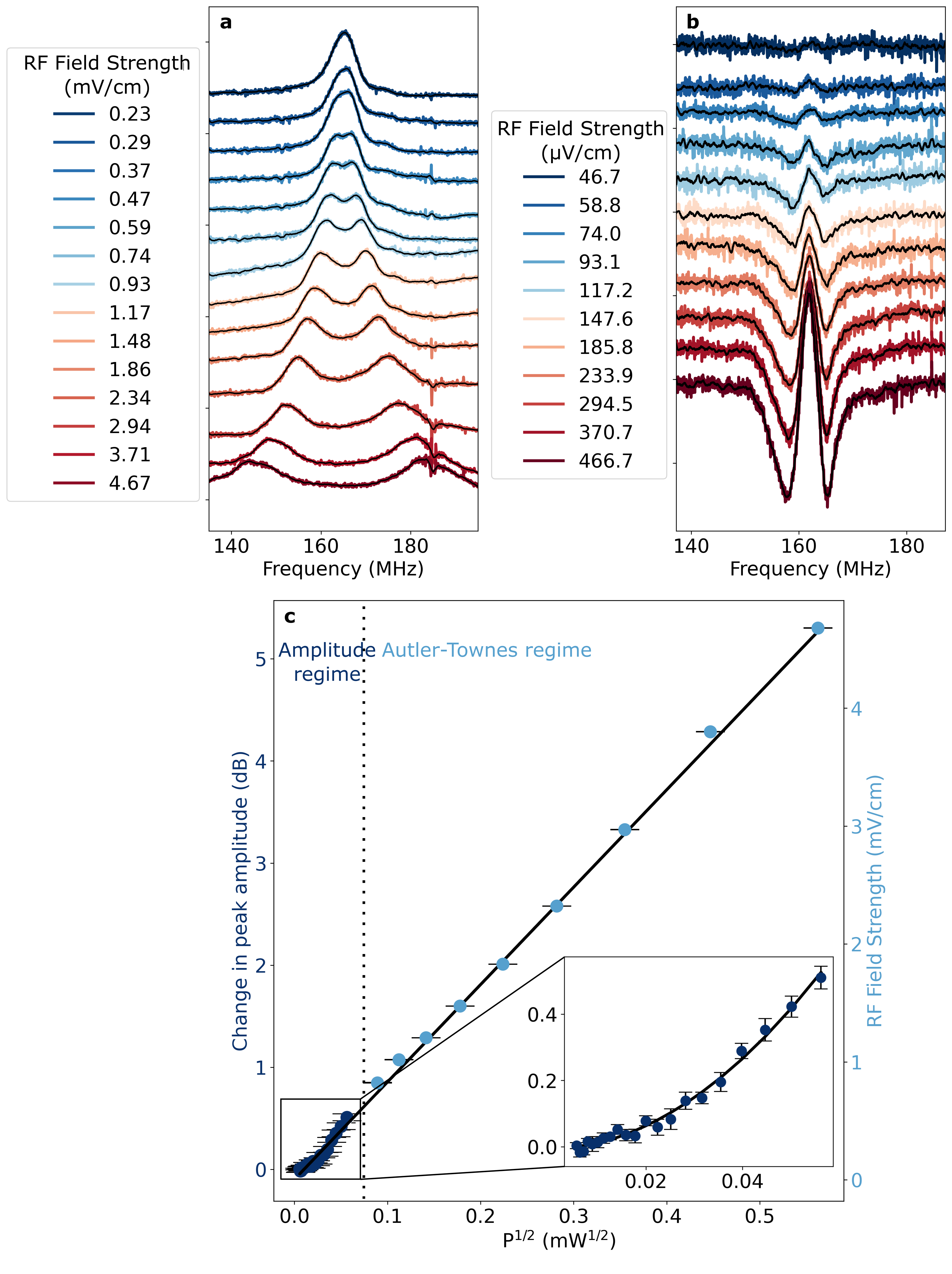}
\caption{(a) RF induced peak splitting in the Autler-Townes regime. (b) Peak amplitude difference induced by RF in the amplitude regime. (c) Peak splitting and amplitude change as a function of $\sqrt{P}$ for both Autler-Townes and amplitude regime. A magnified view of the amplitude regime data with a quadratic fit is shown in the inset.}
\label{fig:splitting}
\end{figure}

The RF electromagnetic field induces Autler-Townes splitting of the 55D$_{5/2}$ peak that produces two transmission windows separated in frequency by,
\begin{equation} \label{eq:rabi}
      \Omega_{RF}=\frac{d\cdot E}{\hbar},
\end{equation}
where the dipole moment of the transition is $d=6294ea_o$ \cite{sedlacek2012microwave}. The splitting of the EIT peak is shown as a function of RF electric field strength in Figure~\ref{fig:splitting}a, as calculated by Equation~\ref{eq:rabi}, wherein the magnitude of the splitting increases with the strength of the RF electromagnetic field. When the probe laser is scanned, a factor of $\lambda_p/\lambda_c$ modifies the spectral splitting due to the Doppler effect. To account for the slope in the comb profile shown in Figure~\ref{fig:comb}a the split peaks are levelled by subtracting a linear background fit.
As the power output of the horn antenna is related to the electric field strength as $P \propto E^2$, we expect that the square root of the RF electromagnetic field power, $P$, should have a linear relationship to the spectral splitting of the EIT feature, which is demonstrated in Figure~\ref{fig:splitting}c. 


At low RF electromagnetic field strengths the magnitude of the EIT peak splitting cannot be measured directly, but results in an amplitude reduction of the transmission peak. The changes in peak amplitudes are shown for low RF electromagnetic field powers in Figure~\ref{fig:splitting}b. The difference in peak amplitude, as determined by averaging the peak amplitude over a 0.4$\,$MHz span centered at 162$\,$MHz, becomes quadratic in the amplitude regime \cite{sedlacek2012microwave}, as shown in the inset of Figure~\ref{fig:splitting}c. The weakest detectable field, defined as the last measurement before the change in peak amplitude drops below zero, is 66$\pm$0.4$\,$$\mu$V$\,$cm$^{-1}$.

\begin{figure}[htbp]
\centering
\includegraphics[width=\textwidth]{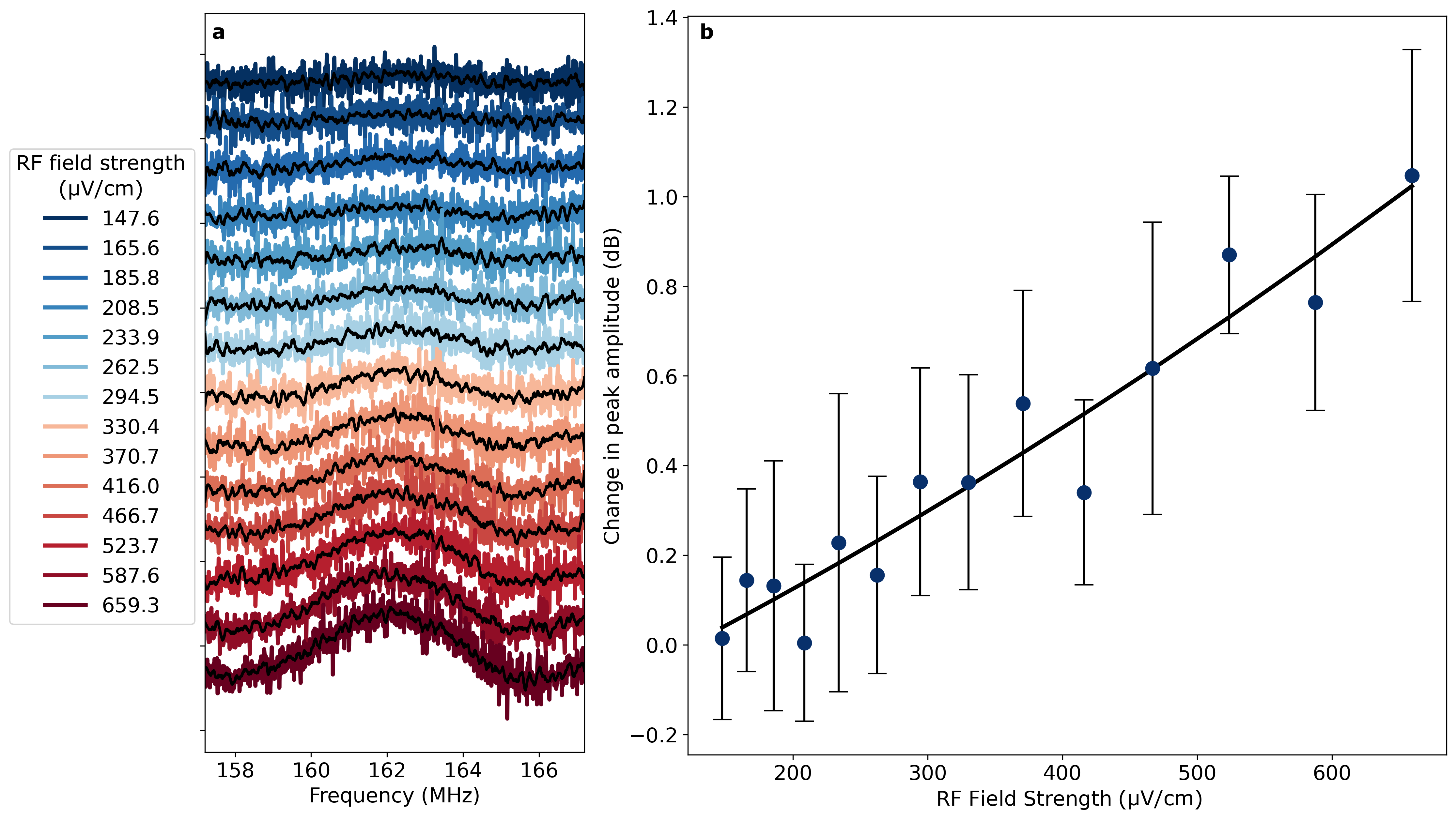}
\caption{(a) Peak amplitude difference induced by the RF electromagnetic field using a 0.1$\,$ms acquisition time (1$\,$ms scan, 5 averaged scans, 100$\,$kHz range). (b) The change in peak amplitude as determined by a Gaussian fit as a function of $\sqrt{P}$. The black curve is a quadratic fit to the data.}
\label{fig:sensitivity}
\end{figure}

To determine the sensitivity limit of the system and how fast data can be obtained, the acquisition time is reduced. The spectrum analyzer scan time is decreased to 1$\,$ms and the scan range is decreased to 10 MHz. Five scans are averaged for each point. The bandwidth of the spectrum analyzer is increased to 5$\,$kHz, decreasing the signal-to-noise ratio. To obtain the sensitivity in a way where it is possible to compare to previous studies \cite{fan2015atom}, we reduced the sampling range to 0.1$\,$MHz during post-processing. Combined, these changes reduce the effective scan time per measurement from 20$\,$s to 100$\,\mu s$, while still allowing for the observation of the full EIT transmission peak. Figure~\ref{fig:sensitivity}a shows the RF electromagnetic field induced change in the transmission peak amplitude obtained using these measurement settings as a function of RF electromagnetic field strength. The difference in peak amplitude is shown in Figure~\ref{fig:sensitivity}b, and demonstrates a minimum detectable RF electromagnetic field strength of 234$\pm$1.2$\,\mu$V$\,$cm$^{-1}$ and a sensitivity of 2.3$\pm0.02$ $\,\mu$V$\,$cm$^{-1}$Hz$^{-1/2}$. The probe laser comb yields comparable sensitivity and accuracy when compared to single-frequency experiments. The recovery of the full EIT lineshape allows for small frequency fluctuations in the peak center to be observed and used to correct the measurement. Observing the full spectrum can be used to reduce noise compared to single-frequency amplitude regime measurements.

As the fundamental time resolution limit for frequency comb spectroscopy is $\Delta t\Delta f \geq 1$ \cite{hebert2016real}, there is potential to further reduce the measurement time and increase sensitivity through the use of an increased tooth spacing and fast electronics. Using a real-time spectrum analyzer is possible so that the spectrum can be acquired without any scanning. A real-time spectrum analyzer can improve the sensitivity since the entire spectral lineshape can be recovered within the same acquisition time bandwidth, reducing the overall uncertainty in the peak height. Although the sensitivity is not state-of-the-art, it is close to that for experiments that do not use an auxiliary RF field. Better detectors and signal processing can improve the sensitivity. It is also possible to use other multi-photon schemes like the co-linear three-photon scheme described in \cite{shaffer2021rydberg}. We do not see any practical limitations to the technique since Rydberg atom lifetimes are on the order of $\mu s$. The probe laser requires little optical power and minimal RF power is required to produce many comb lines over a sufficient spectral bandwidth.

\section*{Conclusion}

We have demonstrated the use of frequency comb spectroscopy for the detection of RF electromagnetic fields compatible with Rydberg atom-based sensing. Using a self-heterodyne frequency comb interference spectroscopy scheme, we generated a probe laser field that is a flat, dense optical comb generated by an AWG and EOM. We achieved EIT detection with and without laser locking, illustrating the versatility provided by the frequency comb. Additionally, we detected a 19.4 GHz RF electromagnetic field in both the Autler-Townes and amplitude regimes with field strengths as low as 66$\pm$0.4$\,\mu$V$\,$cm$^{-1}$ and sensitivities of 2.3$\pm0.02$$\,$$\mu$V$\,$cm$^{-1}$Hz$^{-1/2}$. The comb approach for the probe laser provides advantages over other Rydberg atom-based sensor readouts such as the detection of RF pulse amplitudes. Future improvements, like better detectors and real time spectrum analyzer read-out, promise to significantly advance the utility of the method. The probe laser comb yielded sensitivities comparable to current experiments, particularly those that do not use an auxiliary RF field for signal processing.

Rydberg atom-based sensors use optical preparation and readout to detect RF electromagnetic fields. EIT is a suitable method for the optical preparation of the atoms since it is sub-Doppler and the the lasers required for Rydberg state excitation are readily available. Using a frequency comb as the probe laser for the EIT scheme can introduce important, new advantages to Rydberg atom-based sensors. Laser tuning is no longer required and pulse amplitude detection is facilitated. Although the use of a frequency comb complicates the RF and electro-optic circuitry of a device, these technologies are well-developed. Real-time readout is possible using real-time spectrum analyzers that can be implemented with digital signal processing, i.e., a fast Fourier transform of the time domain signal. The use of electro-optic frequency combs makes it possible to tailor the comb spacing to a particular task. It may be possible in the future to use Kerr type combs so that entire devices can be integrated using photonic integrated circuits. Photonic integrated circuits can include the lasers, modulators, beam combiners and other optical components in one package.

\pagebreak

\section*{Figures}

\pagebreak

\pagebreak

\pagebreak

\pagebreak

\clearpage
\bibliography{biblio.bib}
\bibliographystyle{unsrt}

\end{document}